\documentclass[graybox]{svmult}

\usepackage{mathptmx}       		
\usepackage{helvet}         			
\usepackage{courier}        			
\usepackage{type1cm}       	 	
\usepackage{makeidx}         		
\usepackage{graphicx}        		
\usepackage{multicol}        		
\usepackage[bottom]{footmisc}		

\makeindex             				

\begin{document}

\title*{Petascale computations for Large-scale Atomic and Molecular collisions}
\author{Brendan M McLaughlin and Connor P Ballance}
\institute{Brendan M McLaughlin\at Centre for Theoretical Atomic, Molecular and Optical Physics (CTAMOP), 
							School of Mathematics \& Physics, The David Bates Building,
       							Queen's University, 7 College Park,  Belfast BT7 1NN, UK, \email{b.mclaughlin@qub.ac.uk}
\and Connor P Ballance \at 		Department of Physics,  206 Allison Laboratory,
                            					Auburn University, Auburn, AL 36849, USA \email{ballance@physics.auburn.edu}}
%
%
\maketitle

\abstract*{Petaflop architectures are currently 
		being utilized efficiently to perform large 
		scale computations in Atomic, Molecular 
		and Optical Collisions. We solve the Schr\"odinger or Dirac equation 
		for the appropriate collision problem using the 
		R-matrix or R-matrix with pseudo-states approach.
		We briefly outline the parallel methodology used and 
		implemented for the current suite of Breit-Pauli and DARC codes. Various 
		examples are shown of our theoretical results compared with 
		those obtained from Synchrotron Radiation facilities and from Satellite observations.
		We also indicate future directions and implementation 
		of the R-matrix codes on emerging GPU architectures.}

\abstract{Petaflop architectures are currently 
		being utilized efficiently to perform large 
		scale computations in Atomic, Molecular 
		and Optical Collisions. We solve the Schr\"odinger or Dirac equation 
		for the appropriate collision problem using the 
		R-matrix or R-matrix with pseudo-states approach.
		We briefly outline the parallel methodology used and 
		implemented for the current suite of Breit-Pauli and DARC codes. Various 
		examples are shown of our theoretical results compared with 
		those obtained from Synchrotron Radiation facilities and from Satellite observations.
		We also indicate future directions and implementation 
		of the R-matrix codes on emerging GPU architectures.}

\section{Introduction}
\label{subsec:1}
Our research efforts continue to focus on the development of computational 
methods to solve the Schr\"odinger and Dirac equations for atomic and 
molecular collision processes. Access to leadership-class computers allows 
us to benchmark our theoretical solutions against dedicated collision 
experiments at synchrotron facilities such as the Advanced Light Source (ALS), 
Astrid II, BESSY II, SOLEIL and Petra III and to provide atomic and molecular 
data for ongoing research in laboratory and astrophysical plasma science. 
In order to have direct comparisons with experiment, semi-relativistic or 
fully relativistic computations, involving a large number of target-coupled 
states are required to achieve spectroscopic accuracy. These computations 
could not be even attempted without access to HPC resources such 
as those available at leadership computational centers in Europe and the USA.
We use the R-matrix, R-matrix with pseudo-states (RMPS) method 
to solve the Schr\"odinger and Dirac equations for atomic and molecular collision processes. 

Satellites such as {\it Chandra} and  {\it XMM-Newton}
are currently providing a wealth of x-ray spectra on many
astronomical objects, but a serious lack
of adequate atomic data, particularly in the {\it K}-shell energy range,
impedes the interpretation of these spectra.
Spectroscopy in the soft x-ray region (0.5--4.5~nm), including
{\it K}-shell transitions of singly and multiply charged ionic
forms of atomic elements such as Be, B, C, N, O, Ne, S and Si, as well as L-shell
transitions of Fe and Ni, provides a valuable probe of the extreme
environments in astrophysical sources such as active galactic nuclei (AGN's),
x-ray binary systems, and cataclysmic variables
\cite{McLaughlin2001,Kallman2010,McLaughlin2013}. For example,
{\it K}-shell photoabsorption cross sections for the carbon isonuclear sequence
have been used to model the Chandra X-ray absorption spectrum of the bright blazar Mkn 421 \cite{McLaughlin2010}.

The motivation for our work is multi-fold; (a) Astrophysical Applications \cite{Phaneuf2011}, 
(b) Fusion and plasma modelling, JET, ITER, (c) Fundamental interest and (d) Support of 
experimental measurements and Satellite observations. 
In the case of heavy atomic systems \cite{Ballance2012,McLaughlin2012},
little atomic data exists and our work  provides results for new frontiers on the application of the 
R-matrix; Breit-Pauli and DARC parallel suite of codes.
The current state of parallelism for these codes is outlined and some indication of new directions being 
explored with emerging architectures is presented. These highly efficient codes are widely applicable to
the support of present experiments being performed at  synchrotron radiations facilities 
such as;  ALS, ASTRID II, SOLEIL, PETRA III, BESSY II.
Various examples of large scale calculations are presented to illustrate the predictive nature of the method.

The main question asked of any method is, how do we deal with the many body problem? In our case
we use first principle methods (ab initio) to solve our dynamical equations of motion.  
Ab initio methods provides highly accurate, reliable atomic and 
molecular data (using state-of-the-art techniques) for solving the Schroedinger  and Dirac equation.
The R-matrix non-perturbative method is used to model accurately a wide variety of 
atomic, molecular and optical processes such as; electron impact ionization(EII), 
electron impact excitation (EIE), single and double photoionization and inner-shell X-ray processes.
The R-matrix method provides highly  accurate cross sections and rates  used as input 
for astrophysical modelling codes such as; CLOUDY, CHIANTI, AtomDB, XSTAR 
that are necessary for interpreting experiment/satellite observations of astrophysical objects and 
fusion and plasma modeling for JET and ITER.  

After the Iron Project one needs  to proceed with the Trans-Iron Peak Elements;  (Trans-Iron Peak Project: TIPP),
since various UV spectra of hot post-AGB stars exhibit unidentified lines.
Some of these lines may stem from highly ionized (VI-VIII) metals that are enriched 
by the $s$-process from atomic elements such as: Ge, Ba, Sr, Y, Pb. 
Determination of abundances would be most useful for these elements.
However lines from such ions, are hitherto not identified as the main problem is  precise
 wavelengths are required for unambiguous line identifications.
Some examples are, Ge, Pb, $\dots$, in cooler white dwarfs and sub-dwarfs, 
however abundances are ÒuselessÓ in these cases because they are strongly affected by diffusion.
Palmeri and co-workers have initiated the DESIRE project (http://w3.umons.ac.be/astro/desire.shtml) 
 to address the current deficiency of atomic data for such heavy atomic elements.

\section{Parallel R-matrix Photoionization}
\label{subsec:1}
The use of massively parallel architectures allows one to attempt calculations which previously could not have been addressed.
This  approach enables large scale relativistic calculations for trans-iron elements such as ; Kr-ions, Xe-ions \cite{Ballance2012}.
It allows one to provide atomic data in the absence of experiment and takes advantage of the Linear algebra libraries available 
on most architectures.  We fill in our Òsea of ignoranceÓ i.e. provide data on atomic elements where none have previously existed.  
The present approach has the capability to cater for Hamiltonian matrices  in excess of 250 K x 250 K.
Examples will be shown in the following sections for both valence and inner-shell photoionization for systems 
of prime interest to astrophysics and for more complex species necessary for plasma modelling in fusion tokamaks.

The development of the dipole codes, benefit from similar modifications and developments made to the existing excitation R-matrix codes.
In this case all the eigenvectors from a pair of dipole allowed symmetries are required for bound-free dipole matrix formation.
Every dipole matrix pair  is carried out concurrently with groups of processors assigned to an individual dipole.
The method is applicable to photoionization, dielectronic-recombination or radiation damped excitation and 
now reduces to the time taken for a single dipole formation.  The method so far implemented on various  parallel architectures 
has the capacity to cater for photoionization calculations involving 500 - 1000 levels.  This dramatically improves 
(a) the residual ion structure, (b) ionization potential, (c) resonance structure and  (d) can deal with in excess of 4,000 channels. 

 \section{Scalability}
\label{subsec:1}
As regards to the scalability of these R-matrix codes, we find from experience on a variety of 
petaflop machines that various modules within this suite of codes scale very well, upwards to 100,000 cores.
In practical calculations for cross sections on various systems  it is necessary to perform fine energy resolution of resonance 
features ( ~10$^{-8}$ (Ry) $\sim$ 1.36 meV) which is observed in photoionization cross sections. 
This involves many (6 - 30 million) incident photon energies, vital when comparing with  high Ð precision measurements such 
as those performed on Xe$^{+}$ at the Advanced Light Source synchrotron radiation facility in Berkeley, California, USA 
where energy resolutions of 4 - 9 meV FWHM are achieved.

The formation of many real symmetric matrices (Hamiltonians),
typically 60 K -150 K, requires anywhere from 10-500 Gb of storage.
The diagonalization of each matrix, from which {\it every} eigenvalue
and {\it every}  eigenvector is required is achieved through use of the ScaLapack package.  
In particular routines : {\bf pdsyevx} and {\bf pdsyevd}, where preference is given to the latter, 
as it ensures orthogonality between all eigenvectors. In typical collision calculations 
matrices vary in size from 2K $\times$ 2K to 200K $\times$ 200 K , depending
on the complexity of the atomic target.
The formation of the continuum-continuum part of the N+1 electron Hamiltonian is the most time consuming. Therefore 
if there are several thousand scattering channels then there are [{\it nchan} $\times$({\it nchan} +1)/2] matrix blocks. 
Each block represents a partial wave and each subgroup reads a single Hamiltonian and diagonalizes it in parallel,
concurrently with each other. So there is endless scalability.  R-matrix close-coupling calculations are therefore 
reduced to the time required for a single partial wave.  

In Table 1 we show details of test runs for the outer region module PSTBF0DAMP for 
K-shell Photoionization of B II using 249-coupled states with 400 coupled channels and for  409,600 
energy points and an increasing number of CPU cores. A factor of 4 speed up is achieved by using up to 8192 cores.  
The computations were carried out on the Cray-XE6 (Hopper) at NERSC.  Note, for actual production runs, 
timings would be a factor of 10 larger, as one would require a mesh of 4,096,000 
energy points to fully resolve the resonances features observed in the spectrum.
We present the timings for core sizes varying from 1024 to 8192 again for B II K-shell photoionization in its ground state. 
The computations were done with the outer region module PSTGBF0DAMP for 249-states and 400-coupled channels. 

\begin{table}
\caption{B II 249-states, 400 coupled channels, 409,600 energy points running on increasing number of cores.
		The results are from module PSTGB0FDAMP  for photoionization cross-section calculations of the B II ion carried 
		out on HOPPER the Cray XE6 at NERSC.  Results are presented indicating the speed up factor 
		with increasing number of  CPU cores and the total number of core hours.}
\label{tab-photo}       
%
\begin{tabular}{p{2.8cm}p{2.8cm}p{2.8cm}p{2.8cm}}
\hline\noalign{\smallskip}
CRAY-XE6			&PSTGB0FDAMP 		&PSTGB0FDAMP  	&PSTGB0FDAMP  \\
CPU cores			& Absolute timing (s)		& Speed Up Factor	& Total Core hours\\
\noalign{\smallskip}\hline\noalign{\smallskip}
1024					&584.19				& 1.0000		 	& 166.1155 \\
2048					&430.80				& 1.3584			& 245.0077 \\
4096					&223.08				& 2.6183			&253.8154\\
8192					&149.70				& 3.9018			&340.6506 \\
\noalign{\smallskip}\hline\noalign{\smallskip}
\end{tabular}
\end{table}

The main work horse in our linear algebra code is the ScaLAPACK libraries. 
The goals of the ScaLAPACK project are the same as those of LAPACK; Efficiency (to run as fast as possible), 
Scalability (as the problem size grows so do the numbers of processors grow), Realiability (including error bounds), 
Flexibility (so users can construct new routines from well-designed parts) and
 Ease of Use (by making the interface to LAPACK and ScaLAPACK look as similar as possible).
Many of these goals, particularly portability are  aided by developing and promoting standards, 
 especially for low level communication and computation routines.

Parallel I/O issues have been addressed for the photoionization module, 
PSTGBF0DAMP, as several large passing files (ranging from 10 Gb to -150 Gb) 
need to be read during runtime. These files are SEQUENTIAL FORTRAN BINARY  files. 
For large-scale  computations using the module PSTGBF0DAMP, file names have 
a typical size, e.g. Hamiltonian file  : H.DAT  (10-150 Gb), dipole files : D00,D01,D02,DO3, ...., DNN  (2 - 50 Gb).  
Traditionally, all processors read these files together. For a small numbers of processors, 
say less than 500 processors and file sizes ($<$ 5 Gb) both on the Hopper Cray XE6 (NERSC) 
and the Kraken Cray XT5 (National Institute for Computational Science, NICS) architectures, 
the jobs perform well, with the I/O time being  $<$ 10 \% of the total wall clock time. However, 
as the number of processors increases, i.e 5000-6000 processors, the I/O percentage of 
wall clock time is typically 33\% pre-striping.  This I/O issue has been addressed 
using efficient striping of the data files.  The Lustre file system is made up of an underlying 
set of I/O servers and disks called Object Storage Servers (OSSs) and Object Storage 
Targets (OSTs) respectively. A file is said to be striped when read and write 
operations access multiple OST's concurrently. File striping is used to 
increase the I/O performance of our codes writing or reading from multiple OST's 
simultaneously increases dramatically the available I/O bandwidth.  
We have investigated the following I/O issues along with the use of striping files. 

\begin{enumerate}

\item
  The H.DAT file, unlike the D00,D01, $\dots$ files is read sequentially from start to finish 
  by all processors, however by allowing ONLY processor 0 to read, then distribute (i.e. MPI$\_$BCAST ) 
  the information to the remaining processors is a better solution. This technique however cannot 
  be implemented for the dipole (DXX) files, as each processor is reading different parts of the file, 
  and different amounts of information at each read. Table 2 shows the minimal increase 
  in performance for module PSTGBF0DAMP  for the large-scale photoionization 
  cross-section calculations  on Ca II using this approach.

 \item
Each DXX file holds enough information to describe the photoionization process 
from the ground state and ALL excited states of an atom/ion. If the user/scientist 
requires only a small subset of all possible photoionization cross sections, 
(i.e. ground-state and first metastable) then the DXX files could be REDUCED  in size.

\item
 Modification of the directory structure on the LUSTRE filesystem, so that the 
 large files may be efficiently STRIPED increases the I/O performance of our codes.  
 We have carried out various striping investigations in collaboration with a computational scientist consultant Vince Betro, 
 at NICS, to optimize the 'best striping' based on file size.  Furthermore detailed 
 testing has been done on the Cray XE6 at NERSC (Hopper),  the Cray XT5 at NICS (Kraken), 
 and  on the Cray XE6 at HLRS (Hermit), in Stuttgart, Germany. 
 Our findings are as follows.  For file sizes of $<$ 1 Gb, we use the default striping of files 
 (Hopper has a values of 2, Kraken a value of 1, Hermit a value of 4). 
 For file sizes of between 1 Gb - 10 Gb we use small stripling values, 
 typically 20 gives optimal throughput depending on the system load, which is  only apparent 
 at the NICS facility.   In the case of file sizes between 10 Gb - 100 Gb a medium stripling value of 60 appears to be optimal.  
 For files in excess of 100 Gb a striping value of 120 was used.
\end{enumerate}

%
\begin{table}
\caption{Ca II, 513-states, 1815 coupled channels, 6,000 energy points running on 6,000 cores.
		The results are from module PSTGB0FDAMP  for photoionization of Ca II carried 
		out on HOPPER the Cray XE6 at NERSC.}
\label{tab-photo}       
%
\begin{tabular}{p{3.8cm}p{3.8cm}p{3.8cm}}
\hline\noalign{\smallskip}
CRAY-XE6					&PSTGBF0DAMP$^{a}$			& PSTGBF0DAMP$^{b}$ \\
Ca II							& Absolute timings (s)			& Absolute timings (s) \\
\noalign{\smallskip}\hline\noalign{\smallskip}
6000 cores					& 108.2344$^c$				& 116.2098\\
maximum coupled channels		&1815						& 1815	\\
D file size						& 6 Gb						&6Gb	\\
H matrix size					&23073						&23073	\\
\noalign{\smallskip}\hline\noalign{\smallskip}
\end{tabular}
$^a$ with READ/MPI$\_$BCAST\\
$^b$ without READ/MPI$\_$BCAST\\
$^c$ Time in seconds\\
\end{table}

Optimization of the R-matrix codes on a variety of HPC architectures implements several good coding practices.
We use code inlining (which reduces overhead from calling subroutines), loop unrolling, modularization, unit strides, 
vectorization and dynamic array passing. Highly optimized libraries such as; ScaLapack, BLAS and BLACS 
are used extensively in the codes.  On Cray architectures, the R-matrix codes are profiled with performance utilities 
such as  CrayPat and IPM which gives the user an indication of their scalability.  
Standard MPI/FORTRAN 90/95 programming is used on the 
world's leading-edge petaflop high performance supercomputers.
 Serial (legacy) code to parallel R-matrix implementation  is carried out  using a small subset of basic 
MPI commands.

\section{X-ray and Inner-Shell Processes}
To measure the chemical evolution of the universe and then understand 
its ramifications for the formation and evolution of galaxies 
and other structures is a major goal of astrophysics today.
{\it HUT} and {\it IUE} observations 
have been used to study the processing and dredging that 
goes on during the lifetime of a star. 
Gas that is photoionized by a hard radiation field (AGN/quasars, IGM,
cataclysmic variables, X-ray binaries, etc...) can be 
stripped of its inner-shell electrons, and several phenomenon 
then occur which can radically alter the ionization balance of the gas. 
Space-based UV observations of emission and absorption lines from ions 
of C, N, O, S and Si in photoionized sources play an important role in 
addressing many of the astrophysical issues listed above.  

\begin{figure}[t]
\includegraphics[width=\textwidth]{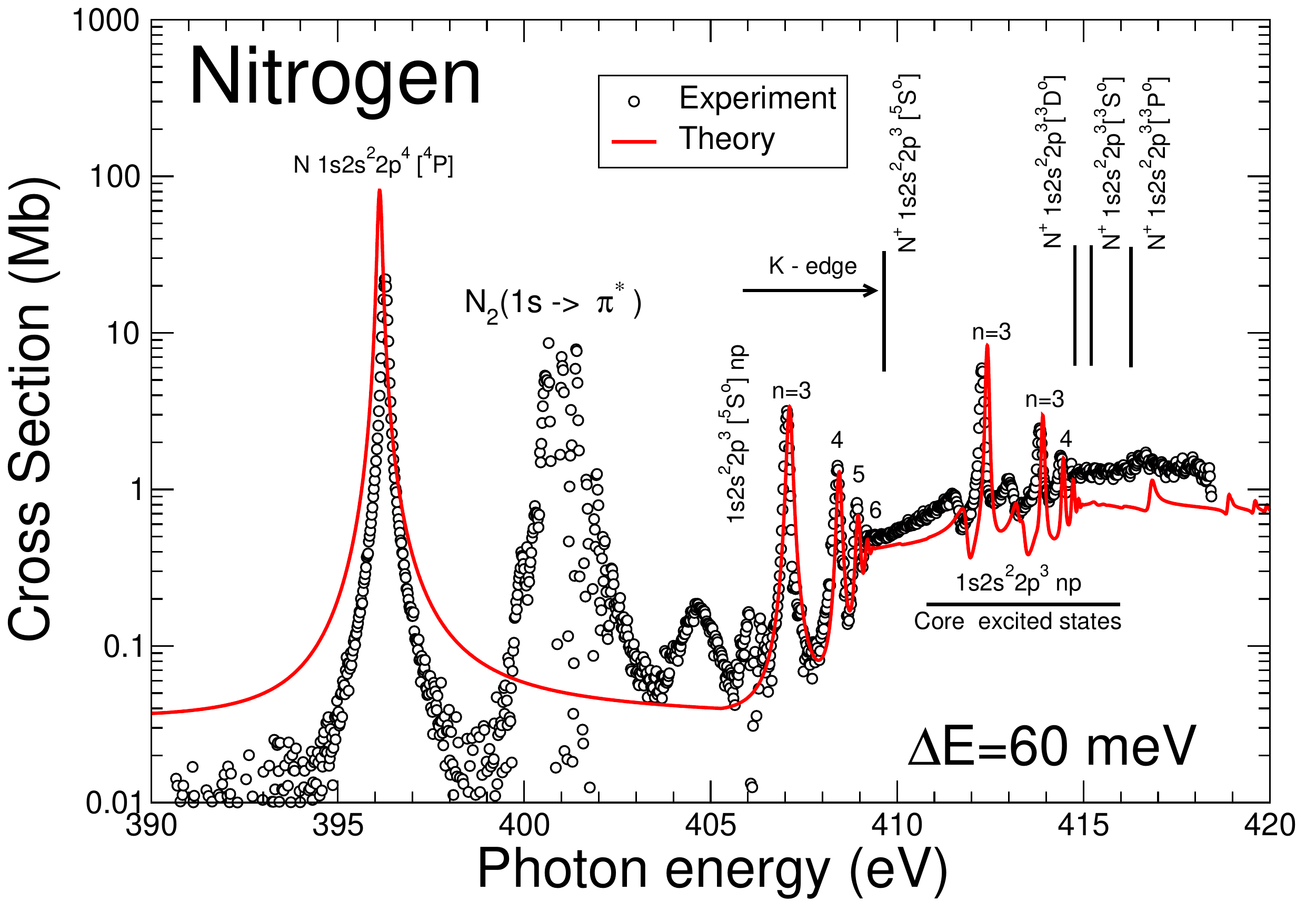}
\caption{\label{nitro}Atomic-nitrogen total photoionization cross section.
                                            Theoretical results from the R-matrix with pseudo-states method (RMPS) were 
                                            convoluted with a 60 meV FWHM Gaussian to simulate the ALS experiment.
				       Experimental results include molecular components between 400 and 406 eV \cite{Marcelo2011}.}
\end{figure}

 Absolute {\it K}-shell photoionization cross sections for atomic nitrogen
have been obtained from high resolution experiment at the ALS 
and the R-matrix with pseudo-states approach (RMPS).  
Due to the difficulty of creating a target of neutral atomic nitrogen,
no high-resolution {\it K}-edge spectroscopy measurements have been reported
for this important atom. Interplay between theory and experiment enabled
identification and characterization of the strong $1s$ $\rightarrow$ $np$ resonance
features throughout the threshold region. An experimental value of 409.64 $\pm$ 0.02 eV
was determined for the {\it K}-shell binding energy. Fig \ref{nitro} illustrates the comparison 
between theory and experiment in the  {\it K}-shell region from our recent work on this system \cite{Marcelo2011}.

\begin{figure}[t]
\begin{center}
\includegraphics[width=\textwidth]{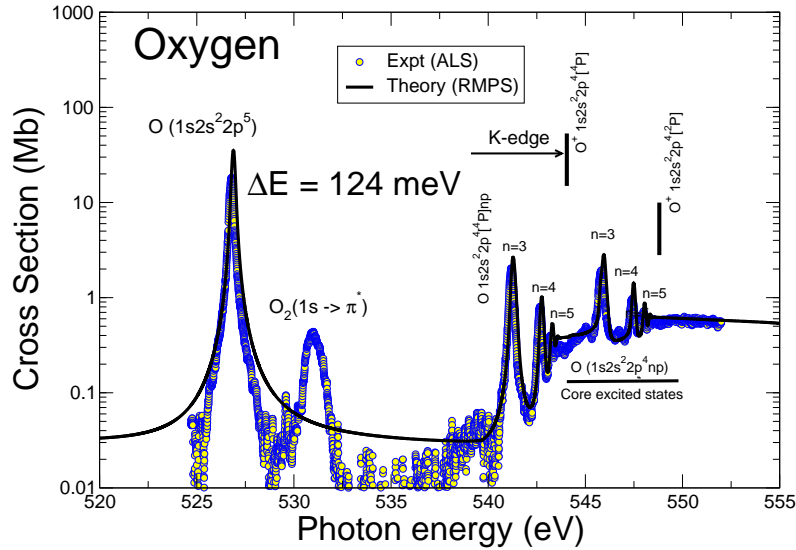}
\caption{\label{figox} (Colour online) Atomic oxygen photo-absorption cross sections 
								taken at 124 meV FWHM  compared with theoretical estimates. 
								The R-matrix calculations shown are  from the
								R-matrix with pseudo-states method (RMPS: solid black line, present results)
								convoluted with a Gaussian
								 profile of 124 meV FWMH \cite{Stolte2013}.}
\end{center}
\end{figure}

An accurate description of the photoionization/photoabsorption of atomic oxygen is important
for a number of atmospheric and astrophysical applications. 
Photo-absorption of atomic oxygen in the energy region below the $\rm 1s^{-1}$ threshold 
 in x-ray spectroscopy from {\it Chandra} and {\it XMM-Newton} is observed in a 
 variety of x-ray binary spectra.  Photo-absorption cross sections 
 determined from an R-matrix method with pseudo-states (RMPS) and high precision measurements 
 from the Advanced Light Source (ALS) are presented in Fig. \ref{figox}.
High-resolution spectroscopy with E/$\Delta$E $\approx$ 4,250 $\pm$ 400 were obtained for
photon energies from 520 eV to 555 eV at an energy resolution of 124 $\pm$  12 meV FWHM. 
{\it K}-shell photoabsorption cross-section measurements were made on atomic oxygen at the ALS.
Natural line widths $\Gamma$ are extracted for the $\rm 1s^{-1}2s^22p^4 (^4P)np~^3P^{\circ}$  and 
$\rm 1s^{-1}2s^22p^4(^2P)np ~^3P^{\circ}$ Rydberg resonances series and compared with theoretical predictions.
Accurate cross sections and line widths are obtained for applications in x-ray astronomy.
Excellent agreement between theory and the ALS measurements is shown which
will have profound implications for the modelling of x-ray spectra and spectral diagnostics. Further details 
can be found in our recent study on this complex \cite{Stolte2013}.

\label{sec:5}

\section{Heavy atomic systems}
\subsection{Kr and Xe ions}
\label{subsec:1}
Photoionization cross sections of heavy atomic elements, in low stages of ionization,
are currently of interest both experimentally and theoretically and for applications in
astrophysics. The data from such processes have many applications in planetary nebulae,
where they are of use in identifying weak emission lines of $n$-capture elements in NGC
3242. For example, the relative abundances of Xe and Kr can be used to determine key
physical characteristics of $s$-process nucleosynthesis, such as the neutron exposure
experienced by Fe-peak seed nuclei.

Xenon and Krypton ions are also of importance in man-made plasmas such as XUV light 
sources for semiconductor lithography, ion thrusters for 
space craft propulsion and nuclear fusion plasmas. 
Xenon and Krypton ions have also been detected in cosmic objects, 
 e.g., in several planetary nebulae  and in the ejected envelopes 
 of low- and intermediate-mass stars \cite{Ballance2012}.

We apply the suite of DARC codes to calculate detailed photoionization (PI) cross sections
on the halogen-like ions, Kr$^{+}$ and Xe$^+$.  
Photoionization (PI)  cross section calculations on the Kr$^+$ 
complex were carried out retaining 326-levels 
in our close-coupling calculations with the Dirac-Atomic-R-matrix-Codes (DARC). 
In R-matrix theory, all photoionization cross section calculations require 
 the generation of atomic orbitals based primarily on 
 the atomic structure of the residual ion.
 The present theoretical work for the photoionization of the Kr$^+$ ion employs
 relativistic atomic orbitals up to n=4 generated for the residual Kr$^{2+}$ ion,
 which were calculated using the extended-optimal-level (EOL) procedure within the GRASP
 structure code\cite{Ballance2012}.

\begin{figure}[t]
\begin{center}
\includegraphics[scale=1.5,width=\textwidth,height=9.0cm]{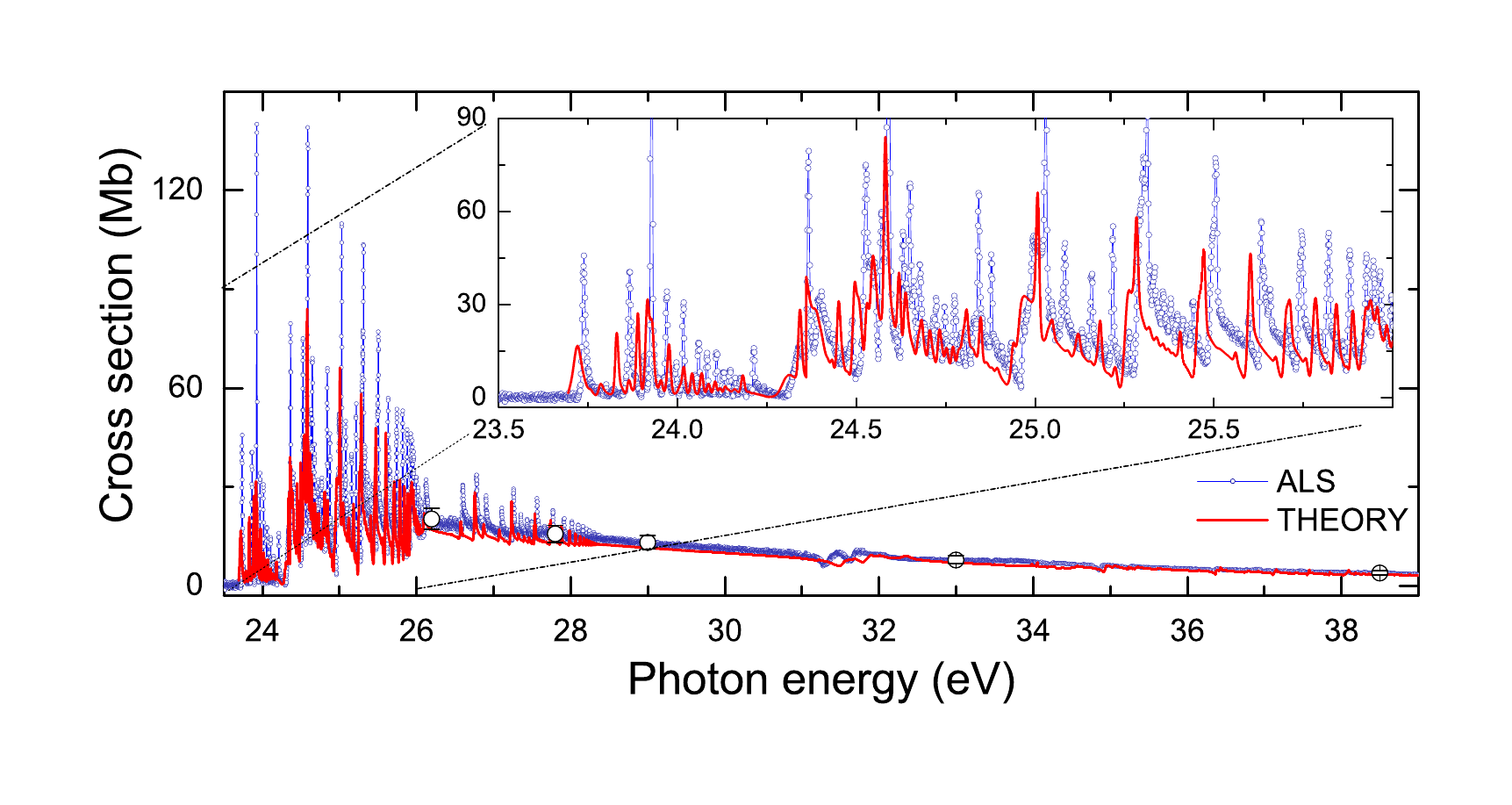}
 \caption{\label{figkr} An overview of measurements for the absolute single photoionization measurements of Kr$^{+}$ ions as a
				function of the photon energy measured with theoretical estimates for cross sections.
				 The ALS measurements are at a nominal energy resolution of 7.5 meV and normalized to the ASTRID/SOLEIL
				  measurements at 26.5025 eV. Theoretical results were obtained from a 
				Dirac-Coulomb R-matrix method, convoluted with the appropriate Gaussian of FWHM
				and a statistical admixture of 2/3 ground and 1/3 metastable states \cite{Hino2012}.}						
\end{center}
\end{figure}
Fig. \ref{figkr} shows the comparison of our large-scale (326-levels) photoionization cross section calculations 
for Kr$^{+}$ ions with the DARC suite of codes compared with the high resolution measurements carried out at the 
Advanced Light Source (ALS), in Berkeley, California, USA.  It is clearly see from Fig. \ref{figkr} 
that our results obtained using the DARC codes reproduce all the fine detail resonance features 
in the observed spectra obtained in the ALS measurements over the photon energy range investigated \cite{Hino2012}.

Similarly photoionization cross section calculations on the Xe$^+$ 
system retained 326 levels of the residual Xe$^{2+}$  ion 
in our close-coupling calculations performed with the Dirac-Atomic-R-matrix-Codes (DARC). 
For the Xe$^+$ case we have employed relativistic n=5 atomic orbitals 
generated for the residual Xe$^{2+}$ ion,
 which were obtained using the energy-average-level (EAL) 
 procedure within the GRASP structure code on the fourteen lowest levels associated 
 with the  $\rm 5s^25p^4$, $\rm 5s5p^5$ and $\rm 5s^25p^35d^2$ configurations.
\begin{figure}[t]
\begin{center}
\includegraphics[width=\textwidth]{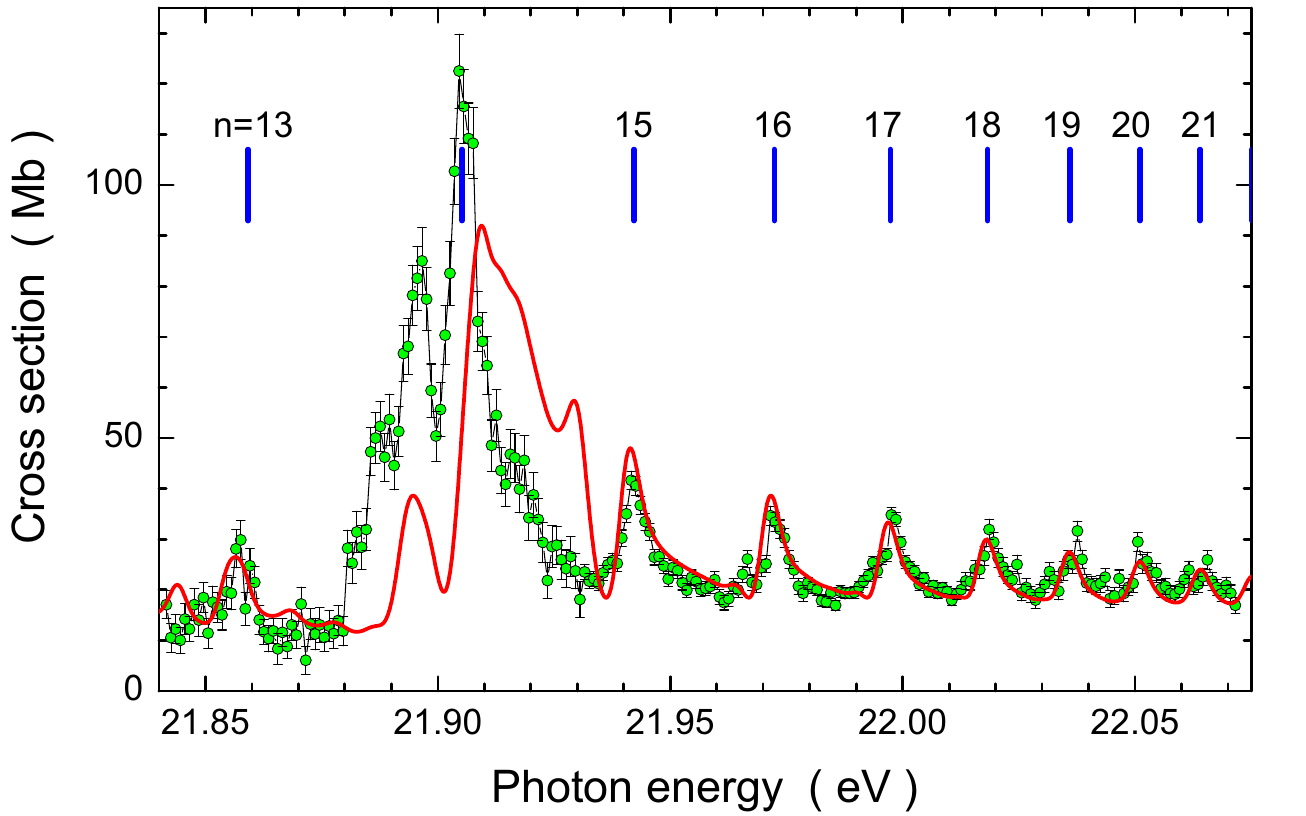}
\caption{\label{figxe} Xe$^+$ ALS experimental PI cross section data (green circles) 
                 for photon energies ranging from 21.84 eV - 22.08 eV at
                a photon energy resolution of 4 meV. Results are compared 
                with theoretical results from a 326-level Dirac-Coulomb R-matrix calculation
               (red line) convoluted with a FWHM Gaussian of 4 meV and statistically 
                averaged over the ground and metastable states to simulate the experimental measurements.
                The bars mark the energies of  the [5s$^2$5p$^4$ ($^3$P$_1$) nd] resonances 
                obtained with a quantum defect of 0.16. \cite{McLaughlin2012}} 
 \end{center}
\end{figure}
For Xe$^+$ ions a more stringent test with recent extremely 
high resolution  experimental measurements made at the ALS at 4 meV as illustrated 
in Fig. \ref{figxe} indicate (in the near threshold region, apart from the n=14 member of the Rydberg series) 
very good agreement giving us confidence in our theoretical results \cite{Ballance2012}.

\subsection{Tungsten (W) Ions}
\label{subsec:2}
The choice of materials for the plasma facing components in fusion experiments is guided by competing 
desirables: on the one hand the material should have a high thermal conductivity, high threshold for melting 
and sputtering, and low erosion rate under plasma contact, and on the other hand as a plasma impurity it 
should not cause excessive radiative energy loss. The default choice of material for present experiments is 
carbon (or graphite), however tritium is easily trapped in carbon-based walls and for that reason carbon is at present 
held to be unacceptable for use in a D-T fusion experiment such as ITER, 
under construction in Cadarache, France, or in a fusion reactor. In its place, 
tungsten (symbol W, atomic number 74) is the favoured material for the wall regions of highest particle and 
heat load in a fusion reactor vessel, with beryllium a possibility for regions of lower heat and particle load. 
ITER is scheduled to start operation with a W-Be-C wall for a brief initial campaign before switching to W-Be or 
W alone for the main D-D and D-T experimental program. In support of ITER and looking ahead to a fusion 
reactor the ASDEX-Upgrade tokamak now operates with an all-W wall, and at JET a full ÒITER-likeÓ mixed 
W-Be-C wall is being installed. Smaller-scale experiments involving tungsten tiles are carried out on other tokamaks. 
The attractiveness of tungsten is due to its high thermal conductivity, its high melting point, and its resistance to 
sputtering and erosion, and is in spite of a severe negative factor that as a high-Z plasma impurity tungsten does 
not become fully stripped of electrons and radiate copiously, so that the tolerable fraction of tungsten impurity 
in the plasma is at most 2$\times$10$^{-5}$. 

The walls of ITER are coated with tungsten which can therefore enter the fusion plasma. W ions impurities in a fusion 
plasma causes critical radiation loss and minuscule concentrations prevent ignition.
High resolution experiments are currently available from the ALS on low ionization stages of W ions.
We use the Dirac-Atomic-R-matrix-Codes (DARC) to perform large scale calculations for the single 
photoinization process and compare our results  with experiment.
These systems are an excellent test bed for the photoionization (PI) process where excellent 
agreement is achieved between theory and experiment providing a road-map for electron - impact excitation (EIE).

\begin{figure}[t]
\begin{center}
\includegraphics[width=\textwidth]{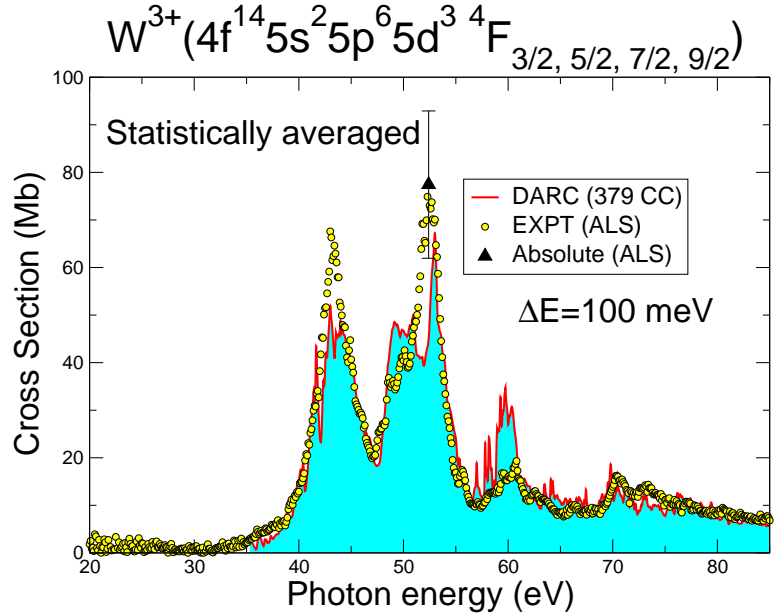}
\caption{\label{w3} Photionization of W$^{3+}$ ions over the photon energy range 20 eV - 90 eV. Theoretical work
				(solid red line: DARC) from the 379 level approximation calculations, convoluted with a Gaussian profile of 
				100 meV FWHM and statistically averaged over the fine structure $J$ =3/2, 5/2, 7/2 and 9/2 levels 
				(see text for details).  The solid circles (yellow) are from the experimental measurements made 
				at the ALS using a band width of 100 meV and the solid triangle (black) is the absolute measurement,
				accurate to within 20\%.}
\end{center}
\end{figure}

For photoionization of the W$^{3+}$ ion of tungsten we investigated several different scattering models before settling on
a  379 Ð level model obtained from 9 configuration state functions of the residual ion.  
Fig. \ref{w3} shows our theoretical results from this 379-level model 
obtained from the DARC codes compared with measurements made at the Advanced Light source. 
From the the comparison made in Fig. \ref{w3} suitable agreement with the ALS experimental 
measurements is obtained when the theoretical results are statistically averaged 
over the fine-structure levels. Additional ions of tungsten are presently being explored and will be reported 
on in due course.

\section{Future directions and emergence of GPUS}
\label{subsec:1}
Experience with the test bed machine DIRAC at NERSC,  a 50 core GPU architecture and 
the Cray XK7 (TITAN) at ORNL has indicated various things.
MPI/Fortran codes access the GPU (Graphical Processing Unit) through the PGI CUDA fortran compiler and 
an architecture supporting NVIDIA graphics cards.  Those using C/C$^{++}$ have greater flexibility.
One of the issues is identifying an intensive numerical algorithm within your code that would benefit 
from distribution across one or more GPUS.  GPUS may be slower than their equivalent CPUS but there are many more cores.
One success has been the formation of the R-matrix itself, which may be expressed 
as a matrix multiplication that has to be carried out repeatedly which occurs in the outer region module PSTGF.
The NVIDIA kepler GPUS on DIRAC outperformed the optimized LAPACK routine DGEMM by a factor of eight, 
though there is  still room to improve the code with CUDA BLAS.
Hence formation of the R-matrix 
\begin{equation}
R_{ij} = \sum_{k}^{~}  \frac{W_{ik} \times W_{kj}}{(E - E_{k})}
\end{equation}
is essentially a matrix multiplication
\begin{equation}
{\bf R} = {\bf X}  \times {\bf Y} 
\end{equation}
where the elements of {\bf R} are ${\it R_{ik}}$,  {\bf X} are ${\it w_{ik}/ (E - E_k)}$ and {\bf Y} are ${\it w_{kj}}$.

%
\begin{table}
\caption{Range of timings (in seconds) using DGEMM for matrix multiplication and the ratio of CPU:GPU 
	       on the Cray XK7 (TITAN) leadership computational architecture located at Oak Ridge National Laboratory. 
	       The results are from module PSTGF and for electron scattering for Fe III.}
\label{tab-gpu}       
%
\begin{tabular}{p{2.8cm}p{2.8cm}p{2.8cm}p{2.8cm}}
\hline\noalign{\smallskip}
Matrix multiplication  			& CPU:GPU		& CPU:GPU	& CPU:GPU \\
$n \times m$					& 1:1 			& 2:1 		& 4:1  \\
\noalign{\smallskip}\svhline\noalign{\smallskip}
~267 $\times$ 16512 			&0.24$^a$  	& 0.35	& 0.30\\
~308 $\times$ 19201			& 0.72 		& 0.52 	& 0.52\\
~399 $\times$  24666			& 0.51  		& 1.10	& 1.12\\
~526 $\times$  32530			& 0.04		 &0.06	& 0.68\\
~132 $\times$ ~8155			&0.22		&0.32	&0.06\\
~260 $\times$16102				&0.90		&1.21	&0.33\\
~837 $\times$51573				&1.03		&1.40	&1.71\\
~796 $\times$49169				&3.43		&1.40	&1.84\\
1368$\times$84321				&3.43		&4.60	& -- \\
1241$\times$76646				&2.73		&3.70	& -- \\
\noalign{\smallskip}\hline\noalign{\smallskip}
\end{tabular}
$^a$ Time in seconds\\
\end{table}

One of the underlying questions is,  how will performance degrade as CPU/GPU ratio increases?  On TITAN, the Oak Ridge 
National Laboratory leadership computational facility, which has a 
hetrogeneous environment (CPU/GPU), a range of timings using DGEMM and the 
ratio of CPU:GPU were investigated for the outer region module, PSTGF.
We investigated the case of electron scattering from the Fe III ion. 
Ten  different matrix sizes were investigated with 16 MPI tasks, 
therefore the MPI process carries out 160 multiplies of each matrix.  
Table \ref{tab-gpu} shows the results for these timings
for  the increasing CPU: GPU ratio and increasing matrix size.

\begin{acknowledgement}
C P Ballance was supported by US Department of Energy (DoE)
grants  through Auburn University. 
B M McLaughlin acknowledges support by the US
National Science Foundation through a grant to ITAMP
at the Harvard-Smithsonian Center for Astrophysics and
a visiting research fellowship from Queen's University Belfast.  
The computational work was carried out at the National Energy Research Scientific
Computing Center in Oakland, CA, USA, the Kraken XT5 facility at the National Institute
for Computational Science (NICS) in Knoxville, TN, USA
and at the High Performance Computing Center Stuttgart (HLRS) of the University of Stuttgart, Stuttgart, Germany.
The Kraken XT5 facility is a resource of the Extreme Science and Engineering Discovery Environment (XSEDE),
which is supported by National Science Foundation grant number OCI-1053575.
This research also used resources of the Oak Ridge Leadership Computing Facility 
at the Oak Ridge National Laboratory, which is supported by the Office of Science 
of the U.S. Department of Energy under Contract No. DE-AC05-00OR22725.
\end{acknowledgement}
%
%
%
%

\end{document}